\documentclass[twocolumn,floatfix,aps,showpacs]{revtex4}
\usepackage{graphicx}
\begin{document}
\title{Random-matrix theory of quantum size effects\\
on nuclear magnetic resonance in metal particles}
\author{C. W. J. Beenakker}
\affiliation{Instituut-Lorentz, University of Leiden,
P.O. Box 9506, 2300 RA Leiden, The Netherlands}
\begin{abstract}
The distribution function of the local density of states is computed exactly
for the Wigner-Dyson ensemble of random Hamiltonians. In the absence of
time-reversal symmetry, precise agreement is obtained with the
``supersymmetry'' theory by Efetov and Prigodin of the NMR lineshape in
disordered metal particles. Upon breaking time-reversal symmetry, the variance
of the Knight shift in the smallest particles is reduced by a universal factor
of 2/3.
\end{abstract}
\date{August 1994}
\maketitle
\narrowtext
A new quantum size effect in small metal particles has been predicted by Efetov
and Prigodin \cite{Efe93}. They computed the spectral lineshape for nuclear
magnetic resonance (NMR) and found that the resonance becomes very broad upon
decreasing the temperature and particle size, due to large fluctuations in the
Knight shift at different points in the sample. Similar results for a
disordered wire have been reported by Mirlin and Fyodorov \cite{Mir94}, who
extended earlier work on one-dimensional conductors
\cite{Yan74,Avi85,Alt89,Vai92}. Recent NMR spectroscopy on a monodisperse array
of nanometer-size Pt particles is in striking agreement with the theory
\cite{Bro94}. The essential difference with quantum size effects on
thermodynamic properties \cite{Hal86} is that NMR in a metal measures the {\em
local\/} density of states $\rho(E,{\bf
r})=\sum_{n}\delta(E-E_{n})|\Psi_{n}({\bf r})|^{2}$, and thus depends both on
the energy levels $E_{n}$ {\em and\/} the wavefunctions $\Psi_{n}({\bf r})$ of
the valence electrons. The sensitivity to the local density of states occurs
because the frequency of microwave absorption by a nucleus is shifted as a
result of the Fermi contact interaction between the nuclear spin and the
electron spin. (This is known as the Knight shift \cite{Sli80}.) The frequency
shift $\delta\omega_{i}$ for a nucleus at position ${\bf r}_{i}$ is linearly
proportional to $\rho(E_{\rm F},{\bf r}_{i})$ (with $E_{\rm F}$ the Fermi
energy). Different nuclei have different frequency shifts, which is observed as
a broadening of the resonance in an ensemble of particles. To determine the
broadening one has to consider the combined effect of particle-to-particle
fluctuations in the energy spectrum and spatial fluctuations of the
wavefunctions within the disordered particles. The strong spectral correlations
make the theory considerably more complex than for an ideal spherical particle
\cite{Yu80,Mak85}.

Efetov and Prigodin computed the fluctuations in the local density of states
from a microscopic model of a disordered metal particle with elastic impurity
scattering in a strong magnetic field, by mapping it onto a supersymmetric
non-linear sigma-model. The traditional approach \cite{Hal86} to quantum size
effects in metal particles is based on random-matrix theory. Following Gor'kov
and Eliashberg \cite{Gor65}, it is assumed that the Hamiltonian ${\cal H}$ of
an irregularly shaped metal particle is a random $N\times N$ Hermitian matrix,
with the Gaussian distribution
\begin{equation}
P({\cal H})=C\exp(-c\,{\rm Tr}\,{\cal H}^{2})\label{Gaussian}
\end{equation}
originally introduced by Wigner and Dyson for the spectrum of a heavy nucleus
\cite{Bro81}. The coefficient $c$ determines the mean level spacing
${\mit\Delta}$ (which in the limit $N\rightarrow\infty$ is uniform in the bulk
of the spectrum), and $C$ is a normalization constant. In the past,
random-matrix theory has been applied to quantum size effects on thermodynamic
properties of metal particles \cite{Hal86,Den71}, in agreement with microscopic
theories \cite{Efe83,Ver85,Alt86}. These applications involve the distribution
of the set of eigenvalues $\{E_{n}\}$ of ${\cal H}$, which follows from
$P(\cal{H})$ on integrating out the eigenvectors $\{\Psi_{n}\}$. In contrast,
the NMR lineshape depends on the joint distribution of the $E$'s and $\Psi$'s.
Problems of this type have not previously been tackled by random-matrix theory.

It is the purpose of this paper to show how the anomalous broadening of the NMR
line shape can be obtained directly from the Wigner-Dyson distribution
(\ref{Gaussian}), without any further assumption. In the absence of
time-reversal symmetry we recover precisely the results of Ref.\ \cite{Efe93}.
Experiments on nanometer-size particles are typically performed in the presence
of time-reversal symmetry. (The authors of Ref.\ \cite{Bro94} estimate that to
break time-reversal symmetry in their system would require magnetic fields of
the order of 1000~T, two orders of magnitude greater than the experimental
fields.) Random-matrix theory is particularly suited to investigate the
dependence of the fluctuations on fundamental symmetries of the Hamiltonian. A
celebrated example is the reduction by a factor of 1/2 of the variance of the
universal conductance fluctuations, upon breaking time-reversal symmetry
\cite{Alt86,Sto91,Bee93}. We will show that the variance of the Knight shift
has a different reduction factor of 2/3, provided the level spacing is much
greater than both the level broadening and the temperature.

Let us first reformulate the problem of the NMR lineshape in the framework of
random-matrix theory. The intensity $I(\omega)$ of the resonance at frequency
$\omega$ is given by the distribution $P(\rho)$ of the local density of states
upon rescaling,
\begin{equation}
I(\omega)=aP\biglb(\rho=b(\omega-\omega_{0})\bigrb),\label{Iomega}
\end{equation}
with microscopic parameters $a,b,\omega_{0}$. The distribution $P(\rho)$ is
defined by
\begin{equation}
P(\rho)=\left\langle\delta\biglb(\rho-\rho(E_{\rm
F},\bf{r})\bigrb)\right\rangle, \label{Prho}
\end{equation}
where the average $\langle\cdots\rangle$ is a spatial average over the total
volume occupied by the particles. The local density of states $\rho(E,{\bf r})$
is given by
\begin{eqnarray}
&&\rho(E,{\bf r})={\textstyle \sum_{n}}f(E-E_{n})|\Psi_{n}({\bf
r})|^{2},\label{rhoEr}\\
&&f(E)=(\gamma/2\pi)(E^{2}+\case{1}{4}\gamma^{2})^{-1},\label{fEgamma}
\end{eqnarray}
where $\gamma$ is the broadening of the levels due to tunneling into the medium
in which the particles are imbedded. Eqs.\ (\ref{Iomega}) and (\ref{Prho})
assume that $\gamma$ is greater than the temperature, the electronic Zeeman
energy, and the spin--orbit scattering rate. For completely isolated particles,
even--odd electron number effects play a role \cite{Hal86}, which are not
considered here. (These effects are expected to be relatively unimportant in
the metal-cluster compounds of current experimental interest \cite{Bro94}.)

For an $N$-dimensional Hamiltonian ${\cal H}$, the continuous variable ${\bf
r}$ is replaced by the index $m=1,2,\ldots N$, and $|\Psi_{n}({\bf r})|^{2}$
becomes $(N/V)|U_{mn}|^{2}$, with $U$ the unitary matrix that diagonalizes
${\cal H}$ and $V$ the volume of a particle. In the absence of time-reversal
symmetry, $U$ varies over the full unitary group. This is relevant to NMR for
very strong magnetic fields and not too small particles, and is the case
considered in Ref.\ \cite{Efe93}. If the flux penetrating a particle is much
less \cite{Dup91} than $h/e$, then time-reversal symmetry is not broken and $U$
is restricted to the orthogonal group. The orthogonal and unitary ensembles are
characterized by the index $\beta=1,2$, which counts how many real numbers
$u_{mn,q}$ ($q=1,\ldots\beta$) define the matrix element $U_{mn}$. There exists
a third symmetry class, characterized by $\beta=4$ and $U$ a symplectic matrix,
which describes systems with time-reversal symmetry in the presence of strong
spin--orbit scattering \cite{Hal86,Bro81}. All our $\beta$-dependent formulas
for the local density of states apply also to the symplectic ensemble, however,
the application to NMR requires a modification of the theory because spin and
charge density are no longer directly related.

For each of the random-matrix ensembles, the average in Eq.\ (\ref{Prho}) can
be written as an integration over eigenvalues and eigenvectors \cite{Meh91},
\begin{eqnarray}
P(\rho)&=&\int\! dE_{1}\cdots\!\int\! dE_{N}\int\! dU\,
C\exp\left(-c\sum_{n}E_{n}^{2}\right)\times\nonumber\\
&&\prod_{i<j}|E_{i}-E_{j}|^{\beta}\,
\delta\biggl(\rho-\frac{N}{V} \sum_{n}f(E_{n})|U_{mn}|^{2}\biggr)
.\label{PrhoUE}
\end{eqnarray}
Here, and in what follows, we choose $E_{\rm F}$ as the zero of energy. The
Jacobian $\prod_{i<j}|E_{i}-E_{j}|^{\beta}$ introduces $\beta$-dependent
correlations between the eigenvalues, in the form of level repulsion. The
eigenvectors are uncorrelated with the eigenvalues, and distributed uniformly
with measure $dU$.

The variance of the Knight shift requires the first two moments of $P(\rho)$.
The general formula for the $p$-th moment is
\begin{equation}
\overline{\rho^{p}}=(N/V)^{p}\left\langle\biglb({\textstyle\sum_{n}}
f(E_{n})|U_{mn}|^{2}\bigrb)^{p}\right\rangle.\label{rhop}
\end{equation}
The first moment evaluates trivially to
\begin{equation}
\bar{\rho}=(V{\mit\Delta})^{-1}\equiv\rho_{0}.\label{rho0def}
\end{equation}
To evaluate the second moment we use the formula \cite{Ull64}
\begin{equation}
\langle|U_{mn}|^{2}|U_{mn'}|^{2}\rangle=\frac{\beta+2\delta_{nn'}}{N(\beta
N+2)}.\label{Umnsquared}
\end{equation}
In the limit $N\rightarrow\infty$, at constant $V$ and ${\mit\Delta}$, we find
\begin{eqnarray}
\overline{\rho^{2}}/\rho_{0}^{2}&=&1+(1+2/\beta){\mit\Delta}\!
\int_{-\infty}^{\infty}\!\!dE\,f^{2}(E)-\nonumber\\
&&{\mit\Delta}^{2}\!\int_{-\infty}^{\infty}\!\!dE
\int_{-\infty}^{\infty}\!\!dE'f(E)f(E')T_{2}(E-E').\label{rho2}
\end{eqnarray}
The two-level cluster function
\begin{equation}
T_{2}(E-E')={\mit\Delta}^{-2}-\langle{\textstyle\sum_{i\neq
j}}\delta(E-E_{i})\delta(E'-E_{j})\rangle\label{T2def}
\end{equation}
is known \cite{Meh91}. For $\beta=2$ one has
\begin{equation}
T_{2}(E)=(\pi E)^{-2}\sin^{2}(\pi E/{\mit\Delta}).\label{T2E}
\end{equation}
The expressions for $\beta=1,4$ are a little more complicated \cite{Meh91}. The
asymptotic behavior of Eq.\ (\ref{rho2}) for ${\mit\Delta}\gg\gamma$ is
obtained from $\lim_{E\rightarrow 0}{\mit\Delta}^{2}T_{2}(E)=1$, hence
\begin{equation}
\overline{\rho^{2}}/\rho_{0}^{2}=(1+2/\beta){\mit\Delta}\!
\int_{-\infty}^{\infty}\!\!dE\,f^{2}(E)+{\cal
O}(\gamma/{\mit\Delta}).\label{asympt1}
\end{equation}
In the opposite regime ${\mit\Delta}\ll\gamma$ one may approximate
${\mit\Delta} T_{2}(E)\simeq\delta(E)$, hence
\begin{equation}
\overline{\rho^{2}}/\rho_{0}^{2}=
1+\frac{2{\mit\Delta}}{\beta}\int_{-\infty}^{\infty}\!\!dE\,f^{2}(E) +{\cal
O}({\mit\Delta}/\gamma)^{2}.\label{asympt2}
\end{equation}

{}From Eq.\ (\ref{rho2}) one readily computes the variance ${\rm
Var}\,K/\bar{K}=\overline{\rho^{2}}/\rho_{0}^{2}-1$ of the Knight shift. The
result is plotted in Fig.\ 1 for $\beta=1,2$ (solid curves). The small and
large-${\mit\Delta}$ asymptotes (dashed) are both linear, but with different
slopes:
\begin{equation}
{\rm Var}\,K/\bar{K}=\frac{2{\mit\Delta}}{\pi\gamma}\times
\left\{\begin{array}{cl} \frac{1}{\beta}&{\rm if}\;\;{\mit\Delta}\ll\gamma,\\
\frac{2+\beta}{2\beta}&{\rm
if}\;\;{\mit\Delta}\gg\gamma.\end{array}\right.\label{varK}
\end{equation}
We have checked that the values for $\beta=2$ agree with Ref.\ \cite{Efe93}.
The transition $\beta=1\rightarrow\beta=2$ on increasing the magnetic field is
signaled by a reduction of ${\rm Var}\,K/\bar{K}$ by a factor of 2/3 for
${\mit\Delta}\gg\gamma$ and 1/2 for ${\mit\Delta}\ll\gamma$. The reduction by
1/2 is the same as for the variance of the universal conductance fluctuations
(UCF) in mesoscopic metals. In that case the broadening of the levels is always
much greater than the level spacing. (Their ratio is the conductance in units
of $e^{2}/h$, which is $\mbox{}\gg 1$ in a metal.) The reduction by 2/3 has no
analogue for UCF.

\begin{figure}[tb]
\centerline{\includegraphics[width=0.9\linewidth]{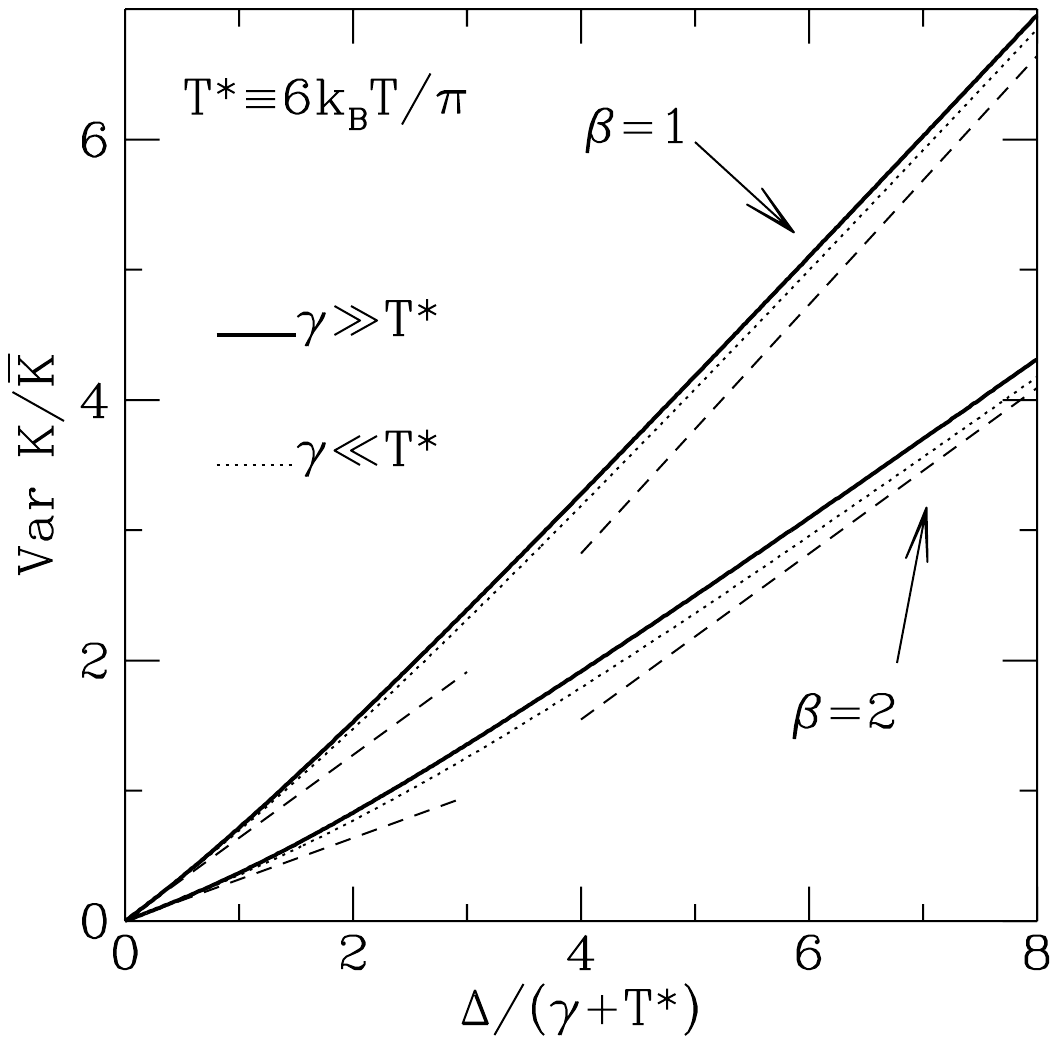}}
\caption{Dependence of the variance of the Knight shift $K$ (normalized by
its average $\bar{K}$) on the level spacing ${\mit\Delta}$, with and without
time-reversal symmetry ($\beta=1$ and 2). The solid curves are computed from
Eq.\ (\protect\ref{rho2}), for the case that the level broadening $\gamma$ is
much greater than the thermal energy $k_{\rm B}T$. The dotted curves are for
the opposite case $\gamma\ll k_{\rm B}T$. The dashed lines are the asymptotes
(\protect\ref{asympt1}) and (\protect\ref{asympt2}) for small and large
${\mit\Delta}$. (The factor $6/\pi$ in the definition of $T^{\ast}$ is chosen
such that the solid and dotted curves have the same asymptotes.) Breaking of
time-reversal symmetry reduces ${\rm Var}\,K$ by a factor of 1/2 and 2/3 for
small and large ${\mit\Delta}$, respectively. These factors are universal,
independent of the relative magnitude of $\gamma$ and $k_{\rm B}T$.
}
\end{figure}

So far we have assumed that the level broadening $\gamma$ is much greater than
the thermal energy $k_{\rm B}T$. At higher temperatures, the function $f(E)$ in
Eq.\ (\ref{rho2}) is to be replaced by the convolution $f_{T}(E)$ of the
Lorentzian (\ref{fEgamma}) with the derivative of the Fermi function. Its
Fourier transform $f_{T}(k)=\int dE\,{\rm e}^{{\rm i}kE}f_{T}(E)$ is
\begin{equation}
f_{T}(k)=\pi k_{\rm B}Tk\biglb[\exp(\case{1}{2}\gamma|k|)\sinh(\pi k_{\rm
B}Tk)\bigrb]^{-1}.\label{fkT}
\end{equation}
The variance of the Knight shift for $\gamma\ll k_{\rm B}T$ is plotted also in
Fig.\ 1 (dotted curves). The asymptotic formulas for small and large
${\mit\Delta}$ become
\begin{eqnarray}
&&{\rm Var}\,K/\bar{K}=\frac{{\mit\Delta}}{k_{\rm B}T}
\,\Phi\!\left(\frac{\gamma}{k_{\rm B}T}\right)\times\left\{\begin{array}{cl}
\frac{1}{\beta}&{\rm if}\;{\mit\Delta}\ll\gamma+k_{\rm B}T,\\
\frac{2+\beta}{2\beta}&{\rm if}\;{\mit\Delta}\gg\gamma+k_{\rm
B}T,\end{array}\right.\nonumber\\
&&\Phi(s)=2\pi\int_{0}^{\infty}\!dq\,{\rm e}^{\textstyle -qs}\, q^{2}(\sinh\pi
q)^{-2}.\label{Fxdef}
\end{eqnarray}
We conclude that the reduction factor associated with
$\beta=1\rightarrow\beta=2$ is universal, independent of the relative magnitude
of temperature and level broadening. This is relevant for experiments, which
are typically in the regime that $k_{\rm B}T$ and $\gamma$ are of comparable
magnitude \cite{Bro94}.

We now turn to the complete distribution $P(\rho)$, given by Eq.\
(\ref{PrhoUE}). It is convenient to work with the (dimensionless) Laplace
transform
\begin{equation}
F(s)=\int_{0}^{\infty}\!\!d\rho\,\exp(-s\rho/\rho_{0})P(\rho),\label{Fsdef}
\end{equation}
and recover $P(\rho)$ at the end by inverting the transform. First, we average
over the eigenvectors. It is known \cite{Bro81,Per83} that, to leading order in
$1/N$, the $\beta N$ components $u_{mn,q}$ ($n=1,2,\ldots N$;
$q=1,\ldots\beta$) of a single row of $U$ are independently distributed
Gaussian variables with zero mean and variance $1/\beta N$. Carrying out the
Gaussian integrations, we find
\begin{eqnarray}
&&F(s)=\left\langle{\textstyle\prod_{n=1}^{N}}
\biglb(1-g(E_{n},s)\bigrb)\right\rangle,\label{Fs1}\\
&&g(E,s)=1-[1+(2s{\mit\Delta}/\beta)f(E)]^{-\beta/2}.\label{gdef}
\end{eqnarray}

The remaining average over the eigenvalues can be carried out using the method
of orthogonal polynomials \cite{Meh91}. This method works for any $\beta$, but
is simplest for the case $\beta=2$. In that case the function $g$ is a
Lorentzian in $E$,
\begin{equation}
g(E,s)=(s\gamma{\mit\Delta}/2\pi)(E^{2}+\case{1}{4}\Gamma^{2})^{-1},
\label{gLorentzian}
\end{equation}
with $\Gamma\equiv\gamma(1+2s{\mit\Delta}/\pi\gamma)^{1/2}$. The large-$N$
limit of Eq.\ (\ref{Fs1}) for $\beta=2$ is given by the Fredholm determinant
\begin{equation}
F(s)={\textstyle\prod_{n=1}^{\infty}}\biglb(1-\lambda_{n}(s)\bigrb),
\label{Fs2}
\end{equation}
where $\lambda_{n}$ is an eigenvalue of the integral equation
\begin{equation}
\int_{-\infty}^{\infty}\!\!dE'\,g(E',s)T_{2}^{1/2}(E-E')
\phi(E')=\lambda\phi(E).\label{FredholmE}
\end{equation}
Fourier transformation gives
\begin{eqnarray}
&&\frac{1}{2\pi}\int_{-\pi/{\mit\Delta}}^
{\pi/{\mit\Delta}}\!dk'\,g(k-k',s)\phi(k')= \lambda\phi(k),\label{Fredholmk}\\
&&g(k,s)=(s\gamma{\mit\Delta}/\Gamma)\exp(-\case{1}{2}\Gamma|k|).\label{gksdef}
\end{eqnarray}
To evaluate the Fredholm determinant of Eq.\ (\ref{Fredholmk}), we discretize
$k\in(-\pi/{\mit\Delta},\pi/{\mit\Delta})$ as
$k_{n}=(\pi/{\mit\Delta})(-1+2n/M)$, $n=1,2,\ldots M$, and then take the limit
$M\rightarrow\infty$ \cite{Note1}:
\begin{eqnarray}
F(s)&=&\lim_{M\rightarrow\infty}{\rm
Det}\,\left[\delta_{nm}-\frac{s\gamma}{M\Gamma}
\exp\left(-\frac{\pi\Gamma|n-m|}{M{\mit\Delta}}\right)\right]\nonumber\\
&=&{\rm e}^{-\alpha\Gamma/\gamma}\,\biglb(\cosh\alpha
+\case{1}{2}(\gamma/\Gamma+\Gamma/\gamma)\sinh\alpha\bigrb). \label{Fs3}
\end{eqnarray}
Inversion of the Laplace transform yields finally
\begin{eqnarray}
P(\rho)&=&\rho_{0}^{-1}(\alpha/2\pi)^{1/2}x^{-3/2}
\exp[-\case{1}{2}\alpha(x+x^{-1})]\nonumber\\
&&\mbox{}\times\biglb(\cosh\alpha+\case{1}{2}(x+x^{-1}-
\alpha^{-1})\sinh\alpha\bigrb).\label{Prhofinal}
\end{eqnarray}
Here $x\equiv\rho/\rho_{0}$ and $\alpha\equiv\pi\gamma/{\mit\Delta}$. Equation
(\ref{Prhofinal}) is precisely the distribution of Efetov and Prigodin
\cite{Efe93}.

This solves completely the problem for $\beta=2$ and zero temperature. For
$\beta=1,4$ and $T\neq 0$ the distribution function can still be written as a
Fredholm determinant, which then has to be evaluated numerically. The
Wigner-Dyson distribution (\ref{Gaussian}) can only describe the pure symmetry
classes ($\beta=1,2$, or 4). The transition between symmetry classes might be
studied by means of an extension known as Dyson's Brownian motion model
\cite{Meh91}. We leave these problems for future work.

In summary, we have derived the result of Efetov and Prigodin \cite{Efe93} for
the NMR lineshape in the absence of time-reversal symmetry from the single
assumption that the Hamiltonian of the particle is a member of the Wigner-Dyson
ensemble of random-matrix theory. A 2/3 reduction of the variance of the Knight
shift in the smallest particles has been predicted to occur upon breaking
time-reversal symmetry.

I am indebted to K. B. Efetov for suggesting this problem as a challenge for
random-matrix theory. B.~Rejaei helped me to compute the determinant in Eq.\
(\ref{Fs3}). Discussions with H. B. Brom on the experimental aspects have been
most helpful. This work was supported by the Dutch Science Foundation NWO/FOM.

\end{document}